\documentclass[11pt,a4paper]{article}
\usepackage{jcappub}

\usepackage{graphics}
\usepackage{bm}

\input epsf

\usepackage{amsmath,amssymb,epsfig,placeins,subfigure,wrapfig,indentfirst}

\begin{document}

\title{Non-singular Brans-Dicke-$\Lambda$ Cosmology}

\author[a]{D.A.Tretyakova,}

\author[b]{A.A.Shatskiy,}

\author[b,c]{I.D.Novikov}

\author[a]{and S.O.Alexeyev}

\affiliation[a]{Sternberg Astronomical Institute, Lomonosov Moscow State University,\\
Universitetsky pr. 13, Moscow, 119991, Russia}

\affiliation[b]{Astro Space Center, Lebedev Physical Institute RAS,\\
84/32 Profsoyuznaya st., Moscow, 117997, Russia}

\affiliation[c]{The Nielse Bohr International Academy, The Nielse Bohr Institute,\\
Blegdamsvej 17, DK-2100 Copenhagen, Denmark}

\emailAdd{tretyakova.d@gmail.com}

\emailAdd{shatskiy@asc.rssi.ru}


\emailAdd{alexeyev@sai.msu.ru}

\abstract{
We discuss a Brans-Dicke model with a cosmological constant, negative value of the $w$ parameter and an arbitrary (in general non-vanishing) scale factor at the Big Bang. The Friedman equations for a flat universe are considered. The current observational values for Hubble constant $H_0$ and deceleration parameter $q_0$ play the role of initial conditions. We follow the approach of \cite{Uehara1982} in order to solve field equations analytically. In Ref. \cite{Uehara1982} only positive values of $w$ were considered, we extend the study to a complete set of possible $w$ values. Our main result is that the scale factor (during it's evolution back in time direction) may not vanish, unlike in the standard $\Lambda CDM$ case. In other words, the considered model demonstrates a cosmological bounce instead of the initial singularity. The famous formula (\ref{Fridman7-5}), that leads to the bounce, is valid only for the dust-filled universe with $p=0$ and, therefore, is not adequate for the Early Universe hot stage when the bounce happens. So, our results are qualitative in nature and must be used to obtain initial values for the hot stage of the Universe.
}


\arxivnumber{1112.3770}

\maketitle

\section{Introduction}
\label{s1}

The cosmic acceleration is now a well observationally established fact  \cite{Riess1998,Riess2004,Perlmutter1999,Schmidt1998,Steinhardt1999,Persic1996}, however it's physical reasons remain open. So, the construction of an adequate cosmological model with a minimal number of ad-hoc hypotheses is a hot topic in modern cosmology. The simplest and hence the most studied model for the moment is cold dark matter one with cosmological constant ($\Lambda CDM$). Providing a good quantitative agreement with observational data, this model however does not explain the nature of the dark matter and dark energy. Another weakness of $\Lambda CDM$ is the absence of explanation of the smallness of $\Lambda$ value if it is assumed to be the so-called ``vacuum energy''. All these arguments lead to the idea of a dynamical theory of dark energy creation (see for example \cite{Copeland2006}).  the most widely discussed candidates are quintessence (a slowly rolling scalar field \cite{Peebles2003}) and higher order curvature gravity (including so-called f(R) gravity models \cite{Buchdahl1970}).

Brans-Dicke model (BD) is one of the first gravity models with a scalar field \cite{BD1961}. It was suggested in 1961 and contains an additional parameter $\omega$ whose value has to be determined by observational data. Large values of $w$ mean an important contribution from the tensor part (Ricci scalar), smaller values of $w$ mean an increasing role for the scalar field contribution. In the limit $w \to \infty$ BD theory leads to General Relativity (GR). In BD model the value of Newton's constant is proportional to the inverse scalar field ($G \approx 1/\phi$), proving additional coupling between the model parameters. The most accurate limit on $w$  comes from Cassini-Huygens mission data on post-Newtonian parameter  $\gamma$ and is $|W| > 50000$ \cite{Bertotti2003}.

BD theory is the most natural GR extension. It is interesting because, firstly, this model could be the low energy effective limit of grand unification (and super unification) approaches (from latest LHC data (see \cite{Wittich2011}) this possibility is not completely closed yet). Second, because the scalar field in BD theory can be reinterpreteted as a dilation fiels in string theory. Finally, because BD model is the simplest GR extension and is useful to investigate any supertheory, so as to gauge the difference with GR \cite{Tamaki1998}. In addition, BD gravity is widely used in cosmology as one needs a scalar field for inflation and such a filed is anyway necessary in the BD model. A large set of inflationary models \cite{Mathiazhagan1984, La1989, Linde1990} is based on BD gravity and more generic scalar-tensor approaches. Brans-Dicke theory is also closely related to the widely discussed f(R) gravity (see for example \cite{Starobinsky2007}).

It is necessary to underline that there is no accelerated expansion in the standard version of the BD model, so, one has to study it's extended versions. One of the most common extensions is a scalar field potential addition. As the accurate shape of this potential is not known yet \cite{Lee2011}, one can consider a $\Lambda$-term as the effective contribution instead of a potential (so we obtain $BD\Lambda$ model). The explanation of the $\Lambda$-term smallness in the framework of  $BD\Lambda$ is possible and was suggested in \cite{Smolyakov2007}. In Ref. \cite{Kim2007}, with the help of scalar field in $BD\Lambda$ context, a dark matter hallo around galaxies is modeled.

An analytical accurate solution for Friedman equations in the $BD\Lambda$ model was also obtained in Ref. \cite{Uehara1982} where positive values of  $\omega$ and initial conditions for the scale factor in the form $a(t_{min})=0$, where $t_{min}$ is Big Bang time, were considered. Partial solutions in this model with scalar field power dependence versus the scale factor were presented in \cite{Ahmadi1995, Pimentel1985, Ram1990, Pandey2001}. Vacuum solutions were obtained in \cite{Romero1993, Romero1992, Cervero1984}. Some papers discussed a $\Lambda$-term dependence as a function of the scalar field (for example, \cite{Etoh1997}). Numerical integration and stability analysis of $BD\Lambda$+matter solutions were carried out in  \cite{Kolitch1996}. Big Rip solution in $BD\Lambda$ is discussed in \cite{Rami2010}.
The analytical solution in the pure BD model with negative $w$, avoiding the initial singularity, was obtained in 1973 by Gurevich et al. \cite{Gurevich1973}. There is no cosmic acceleration in the Gurevich et al. solution, so, nowadays one has to extend BD theory to include cosmic acceleration. In this paper we therefore study the Einstein-Friednman equation solutions in $BD\Lambda$ theory for $w < 0$ with a scale factor with initial value $a(t_{min})=a_{min}$. Generally $a_{min} \neq 0$. Friedman equations are studied for a flat Universe. We take the current values for Hubble parameter and its derivative (deceleration parameter) as their initial conditions. In our study, we use the approach suggested in \cite{Uehara1982}.  We would like to point out that only positive values of $w$ are considered in \cite{Uehara1982}, so, our solution with $w < 0$ represents a new branch. As opposed to the standard $\Lambda CDM$-model, in the considered case,  the scale factor never vanish during backward time evolution. A so-called ``bounce'' (a snap back from the minimal value of $a_{min}$) corresponds to this situation. The expression \ref{Fridman7-5} leading to the bounce is obtained for a cold Universe with $p=0$ and is not valid for a hot Universe. Therefore all the values in the bounce region are only qualitative estimates for the initial values for the transition to the hot stage. 

This paper is organized as follows: \\
in section~\ref{s2} we discuss the choice of the space-time metric and the corresponding field equations;
in section~\ref{s3} the initial values for cosmological parameters are obtained;
in section~\ref{s4} we obtain an analytical solution with a bounce for a dust-filled universe (${p=0}$);
section~\ref{s5} contains a preliminary discussion of the results of section \ref{s4};
in section~\ref{s6} we explore  the case of a ultrarelativistic state of matter (hot phase);
section~\ref{s7} is devoted to the conclusions.

\section{Field equations}
\label{s2}

TThe Friedmann-Robertson-Walker (FRW) metrics reads\footnote{Here and below we set light speed ${c=1}$.}:
\begin{equation}
ds^2 = dt^2 - a(t)^2\left[ \frac{dr^2}{1 - kr^2} + r^2d\Omega^2_2 \right],
\label{1-ds2}
\end{equation}
where ${k=0,\,\pm 1}$.

The action of the BD-$\Lambda$ theory can be written as:
\begin{equation}
S=\frac{1}{16\pi}\int{d^4x\sqrt{-g}\left[\Phi (R + 2\Lambda)
-\frac{w}{\Phi}g^{\mu\nu}\partial_{\mu}\Phi\partial_{\nu}\Phi +16\pi L_{matter}\right]}.
\label{1-S}
\end{equation}
Here $w$ is the BD theory parameter, ${\Phi (t)}$ is the scalar field, $\Lambda$ is the cosmological constant \footnote{$\Lambda$ here can differ from the one in ${\Lambda CDM}$ theory.}.

Variation of the action with respect to the metric $g_{\mu\nu}$ and the scalar field $\Phi$ gives the following field equations:
\begin{eqnarray}
G_{\mu\nu} = \frac{8\pi}{\Phi}T_{\mu\nu} +\Lambda g_{\mu\nu} +\frac{w}{\Phi^2}\left(
\partial_{\mu}\Phi\partial_{\nu}\Phi - \frac{1}{2}g_{\mu\nu} g^{\sigma\lambda}
\partial_{\sigma}\Phi \partial_{\lambda}\Phi \right) +
\frac{\nabla_{\mu}\nabla_{\nu}\Phi-g_{\mu\nu}\nabla_{\lambda}\nabla^{\lambda} \Phi}{\Phi},
\label{eq3} \\
\frac{8\pi}{\Phi}T_{\mu}^{\mu} + 2\Lambda = \frac{3+2w}{\Phi}\nabla_{\lambda}\nabla^{\lambda}\Phi
, \label{eq4}
\end{eqnarray}
where ${\nabla_{\mu}}$ is a covariant derivative,
\begin{eqnarray}
G_{\mu\nu} = R_{\mu\nu}-\frac{1}{2} R g_{\mu\nu} \, ,\quad T_{\mu\nu}=(\rho +p)u_{\mu} u_{\nu} - p
g_{\mu\nu} \, ,\quad
\partial_\mu\Phi = \delta^t_\mu\partial_t\Phi\, .
\end{eqnarray}
Here ${\rho (t)}$ and ${p(t)}$ are the matter density and pressure respectively, the stress-energy tensor corresponds to a barotropic fluid, $G_{\mu\nu}$ is the Einstein tensor.

Here it is convenient to introduce new dimensionless variables \footnote{Here and below the present time is denoted by the subscript "0", so $G_0$ is the the current value of the gravitational constant. From now and further we consider the current time moment as the initial one, so that $t_0=0$. Current values of cosmological parameters are taken as initial conditions. New variables lead to ${\phi_0=1}$, which is convenient for further calculations.}:
\begin{eqnarray}
\Phi (t) \equiv \phi (t)/G_0 , \quad \epsilon(t)\equiv\partial_t\phi/(\sqrt{\Lambda}\phi) ,
\label{1-phi}\\
\tilde H(t)\equiv H(t)/\sqrt{\Lambda}= \partial_t a/(\sqrt{\Lambda}a) , \quad \tilde\rho (t) =4\pi
G_0\rho /\Lambda ,\quad
\tilde p(t) =4\pi G_0 p /\Lambda . \label{1-H} \end{eqnarray}
Here ${H}$ is the Hubble parameter, and ${\tilde H}$ is its dimensionless value.
In these notations, Friedman equations for a flat universe (${k=0}$) in a comoving frame ${(u_\mu =[1,0,0,0])}$ are:
\begin{eqnarray}
\frac{G_t^t}{\Lambda}= 3{\tilde H}^2 = \frac{2\tilde\rho}{\phi} +1+ \frac{w}{2}\epsilon^2 -
3{\tilde H}\epsilon, \label{Fridman1}\\
\frac{G_r^r}{\Lambda}= 2\dot{\tilde H} + 3{\tilde H}^2 = -\frac{2\tilde p}{\phi} +1-
\frac{w}{2}\epsilon^2  -\frac{\ddot\phi}{\phi} - 2{\tilde H}\epsilon . \label{Fridman2-2}
\end{eqnarray}
The Klein-Gordon equation (\ref{eq4}) can be rewritten as:
\begin{equation}
\frac{2\tilde\rho - 6\tilde p}{\phi} +2 = (3+2w)\left[\frac{\ddot\phi}{\phi} + 3{\tilde H}\epsilon\right].
\label{Fridman3}
\end{equation}
Here and below the dot denotes the derivative w.r.t. the dimensionless time ${\tilde t\equiv \sqrt{\Lambda} t}$.

Equations (\ref{Fridman1}-\ref{Fridman3}) lead to the continuity one in the form:
\begin{equation}
\frac{\dot{\tilde\rho}}{\tilde\rho +\tilde p}+3{\tilde H} = 0,
\label{Fridman3-2}
\end{equation}
which is consistent with the equivalence principle.

\section{Initial values of the model parameters}
\label{s3}

We introduce the deceleration parameter $q$ and the dimensionless matter density $\beta$ for the initial time in the following form:
\begin{eqnarray}
\dot{\tilde H}\equiv -(1+q){\tilde H}^2 \, ,\quad
\beta\equiv\frac{4\pi G_0 (\rho_0 - p_0)}{H_0^2} = \frac{\tilde\rho_0 -\tilde p_0}{{\tilde H}_0^2}
\, .
\label{initial-0}
\end{eqnarray}
Combining the equations (\ref{Fridman1}-\ref{Fridman3}) to exclude ${\epsilon}$ and ${\ddot\phi /\phi}$, we obtain for $p=0$ at the current moment $t_0$ the following equation:
\begin{eqnarray}
w \left[ {\tilde H}_0^2(2-q_0-\beta z)-z\right]^2 - 2{\tilde H}_0^2(3z-1) +{\tilde H}_0^4(6-6q_0-6\beta z+4\beta) =0,
\,\,\, z\equiv \frac{2+2w}{3+2w} .
\label{initial-1}\end{eqnarray}
This equation defines $H_0$ as a function of $\beta$, $q_0$ and $\omega$. In the ${|w|>>1}$ approximation\footnote{Here and below (unless otherwise noted) the arrow denotes the ${|w| >> 1}$ approximation.} equation (\ref{initial-1}) yields:
\begin{eqnarray}
\frac{1}{{\tilde H}_0^2} \to (2-q_0-\beta) \pm\sqrt{\frac{2(1+q_0-\beta)}{w}}.
\label{initial-2}
\end{eqnarray}
In the GR limit ($w \to \infty$) the second term can be neglected, so, current cosmological parameters values ban be established as \cite{cai, komatsu}: \\ ${H_0\approx 2.3\cdot 10^{-18}\mbox{sec}^{-1}}$, ${\,\,\rho_0\approx 0.27\cdot 10^{-29}\mbox{g/cm}^3}$ (accounting for baryonic and dark matter), ${\,\, q_0\approx -0.6}$. We consider a dust-filled universe, thus neglecting the pressure.
We rewrite the above expression for the cosmological constant:
\begin{eqnarray}
\Lambda \to (2-q_0)H_0^2 - 4\pi G_0 (\rho_0 -p_0) \approx 11.3\cdot 10^{-36}sec^{-2}
\label{initial-2-2}\end{eqnarray}
From the lunar ranging experiment (LLR) data \cite{williams} one can extract the following limitations: ${|\partial_t G/G|_{(0)} \leq 4 \cdot
10^{-20} \mbox{sec}^{-1}}$, hence ${|\epsilon_0|}$ is a small value: ${|\epsilon_0| < 0.01}$.
When ${|w|>> 1}$ we have:
\begin{eqnarray}
{\tilde H}_0\approx 0.68\, ,\quad \tilde\rho_0\approx 0.2 \, ,\quad \beta\approx 0.4 \, .
\label{initial-3}\end{eqnarray}
Combining (\ref{Fridman1})$+$(\ref{Fridman2-2}), multiplying the result by ${1/{\tilde H}_0^2}$ and substituting ${\ddot\phi /\phi}$ from (\ref{Fridman3}), we obtain for the initial moment $t_0$ the following expression:
\begin{eqnarray}
\frac{\epsilon_0}{{\tilde H}_0} = \frac{1}{{\tilde H}_0^2} - (2-q_0-\beta) + \frac{\beta +1/{\tilde H}_0^2}{3+2w}
\label{initial-4-0}\end{eqnarray}
Substituting the $H$  value from (\ref{initial-2}), we obtain the order of ${1/\sqrt{w}}$\footnote{When calculating the rhs of the expression (\ref{initial-4-0}) we only considered the terms of order ${1/\sqrt{w}}$ from (\ref{initial-2}),  the last term from (\ref{initial-4-0}) was neglected due to the taken accuracy.}:
\begin{eqnarray}
\epsilon_0 \to \pm\sqrt{\frac{2(1+q_0-\beta)}{w(2-q_0 -\beta)}}
\label{initial-4}\end{eqnarray}

\section{Dust-filled Universe solution}
\label{s4}

First of all we consider a dust-filled Universe, i.e. ${p=0}$. As is Ref. \cite{Uehara1982}, we rewrite field equations using ${f\equiv\phi a^3}$, and take into account that the expression (\ref{Fridman3-2}) leads to ${\tilde\rho /\phi =\tilde\rho_0 f_0/f}$.

Considering  ${\ddot f/f = \ddot\phi /\phi +6{\tilde H}\epsilon +3\dot{\tilde H}+9{\tilde
H}^2}$, we combine field equations in the following way:
${\frac{3}{2}}[$(\ref{Fridman1})$+$(\ref{Fridman2-2})$]+$(\ref{Fridman3})$/[6+4w]$. This yields to:
\begin{equation}
\ddot f - \eta^2 \left( f + \tilde\rho_0 f_0\right) =0 \, ,\quad \eta^2\equiv\frac{8+6w}{3+2w} .
\label{Fridman7-1}
\end{equation}
The obtained equation can be straightforwardly integrated:
\begin{equation}
\frac{f(\tilde t)}{f_0}=c^{+}E + c^{-}/E - \tilde\rho_0 \, ,\quad
E(\tilde t) \equiv \exp (\eta\tilde t) \, .
\label{Fridman7-2}
\end{equation}
where ${c^+}$ and ${c^-}$ can be easily obtained from the initial data.

One can rewrite (\ref{Fridman3}) as:
\begin{equation}
2f + 2\tilde\rho_0 f_0 = (3+2w)(\dot\phi a^3\dot{)}.
\label{Fridman7-3}
\end{equation}
With the help of Eq. (\ref{Fridman7-2}) one gets the expression for the Hubble parameter from Eq. (\ref{Fridman7-2})\footnote{Note that ${\,\, d\tilde
t=dE/(\eta E)}$.}:
\begin{eqnarray}
3{\tilde H} = \frac{\dot f}{f} - \frac{\dot\phi}{\phi} =
\frac{\dot f}{f} -\frac{2f_0}{f(3+2w)} \int\limits_{const}^{\tilde t}\left(\frac{f}{f_0}
+\tilde\rho_0 \right)\, d\tilde t =
\frac{\dot f}{f} - \frac{2(c^{+}E -c^{-}/E + c_{_H})}{\eta(3+2w)(c^{+}E + c^{-}/E - \tilde\rho_0)}
=\nonumber\\
= \frac{6(1+w)(c^+ E -c^- /E)-2c_{_H}}{\eta(3+2w)(c^+E + c^- /E -\tilde\rho_0)}. \qquad
\label{Fridman7-4}
\end{eqnarray}
Here ${c_{_H}}$ can also be determined from initial data.

Solving (\ref{Fridman7-2}) and (\ref{Fridman7-4}) for the present time, one obtains the coefficient values as:
\begin{eqnarray}
c^+ = \frac{1+\tilde\rho_0}{2} + \frac{\epsilon_0 +3{\tilde H}_0}{2\eta} \, ,\quad c^- =
\frac{1+\tilde\rho_0}{2} - \frac{\epsilon_0 +3{\tilde H}_0}{2\eta} \, ,\quad c_{_H} =
\frac{\eta\epsilon_0 (3+2w)}{2} - \frac{\epsilon_0 +3{\tilde H}_0}{\eta} \, ,
\label{Fridman7-6}\end{eqnarray}
resulting in the following expression for the scale factor:
\begin{eqnarray}
\frac{a}{a_0} = \left(c^{+}E + c^-/E -\tilde\rho_0\right)^{1/3}
\exp\left[ \frac{-1}{3(4+3w)}\int\limits_1^E \left(
\frac{c^{+}E^2 + c_{_H}E -c^{-}}{c^{+}E^2 - \tilde\rho_0 E + c^{-}}\right) \frac{dE}{E}\right]
=\nonumber \\
= \left(c^{+}E + c^-/E -\tilde\rho_0\right)^{\frac{1+w}{4+3w}}
\exp\left[ \frac{-2c_{_H} (A-A_0)}{3(4+3w)\sqrt{\Delta}} \right],
\label{Fridman7-5}\end{eqnarray}
where
\begin{eqnarray}
\Delta\equiv 4c^+ c^- - {\tilde\rho_0}^2 = 1+2\tilde\rho_0 - (3{\tilde H}_0 + \epsilon_0)^2/\eta^2
= \frac{-3}{8+6w}\left[ {\tilde H}_0 -\epsilon_0 (1+w) \right]^2
\, ,\nonumber\\
A(E)\equiv \arctan [(2c^+ E -\tilde\rho_0)/\sqrt{\Delta}] \,
.\qquad  
\label{Fridman7-7}
\end{eqnarray} 
\begin{eqnarray}
\phi = \left(c^+ E + c^-/E - \tilde{\rho_0} \right)^{\frac{1}{4+3w}} \mbox{exp}\left[ \cfrac{2c_H (A - A_0)}{(4+3w)\sqrt{\Delta}}\right].
\label{Fridman7-7a}
\end{eqnarray} 
In order to keep $\Delta$  positive, it is necessary to set $w$ to be rather large (${|w|>>1}$) and negative\footnote{It is also importantly to mention that the ${w<0}$ case in BD model opens the possibility for wormholes existence without energy conditions violation, see ~\cite{Agnese1994, Rannu2011} for details.}.

In the GR case ${|w|\to\infty}$ and expression (\ref{Fridman7-5}) tends to the well known Friedman solution with acosmological constant:
\begin{eqnarray}
H_{Fr}=\frac{1}{\sqrt{3}}\cdot\frac{E+E_{cr}}{E-E_{cr}}\, ,\quad E_{cr}\equiv\frac{\sqrt{3}{\tilde
H}_0 -1}{\sqrt{3}{\tilde H}_0 +1} \, , \quad
\eta_{Fr} =\sqrt{3} \label{Fridman7-8-1}\, , \\
\frac{a_{Fr}}{a_0} = \frac{(\sqrt{3}{\tilde H}_0 +1)^{2/3} (E-E_{cr})^{2/3}}{(4E)^{1/3}} .
\label{Fridman7-8-2}\end{eqnarray}
It is necessary to note that in this case ${E=E_{cr}}$, ${\Delta =0}$, ${a =0}$, and the scale factor $a(t)$ experiences a kink (which is absent when $\Delta >0$). The Big Bang corresponds to the moment ${t_1\approx -1.46\Lambda^{-1/2}}$, ${\,\, \Lambda^{-1/2}\approx 10^{10}}$ years.

\section{Non-singular cosmology}
\label{s5}

\begin{figure*}
\subfigure{\includegraphics[width=0.49\textwidth]{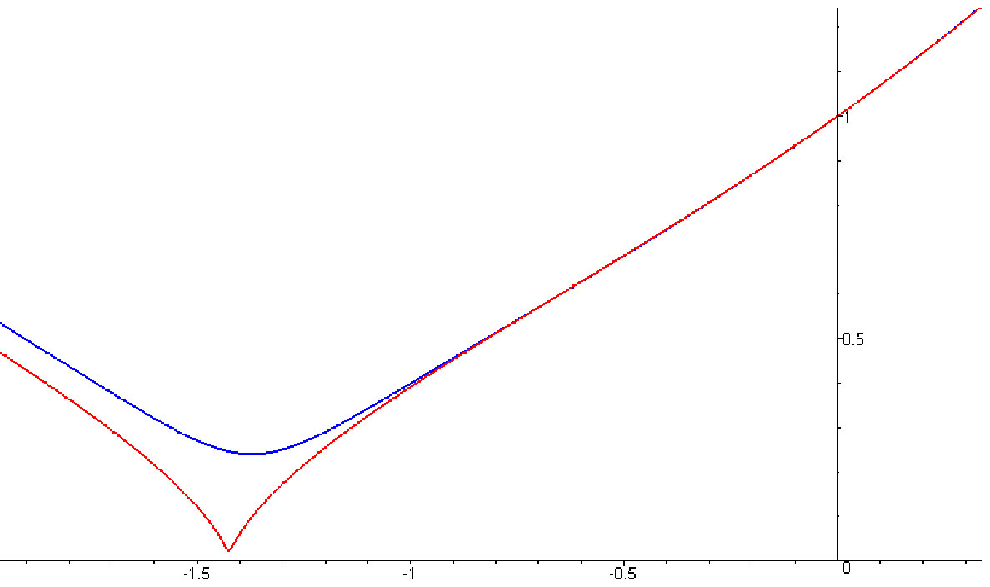}}
\subfigure{\includegraphics[width=0.49\textwidth]{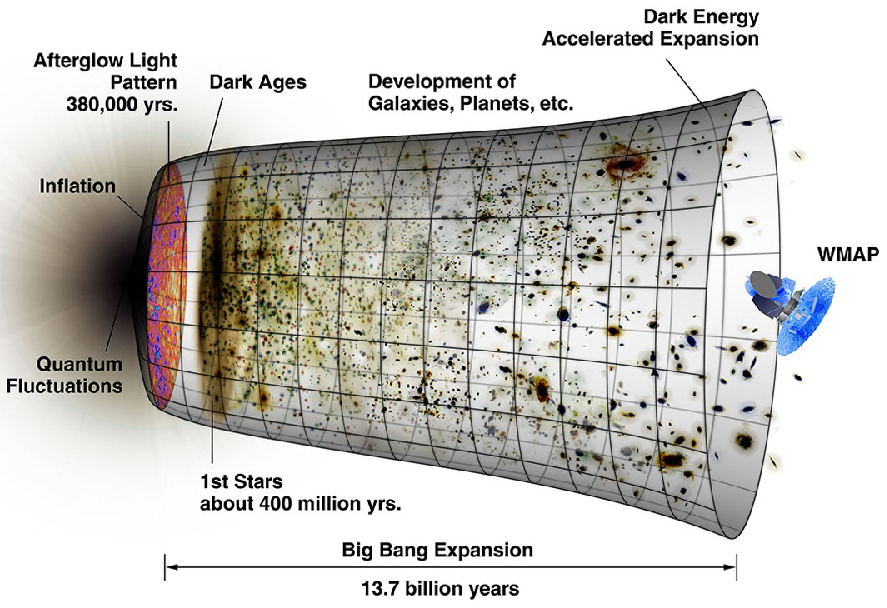}}
\caption{{Left hand side represents ${a(\tilde t)/a_0}$ according
to the expression (\ref{Fridman7-5}) for a dust-filled universe with
a bounce for the following parameter values: ${w=-1000\, ,\,\,\,
q=-0.6}$, ${\beta=0.45}$ for the upper line and ${\beta=0.43653}$
for the lower one. Time unit is ${10^{10}}$~yr. Right
hand side:~illustration (taken from NASA:
http://map.gsfc.nasa.gov) for ${a(\tilde t)/a_0}$ in
$\Lambda$CDM-theory, corresponding to the Friedman solution.}}
\label{R1}
\end{figure*}

In BD$\Lambda$ models, the scale factor may not vanish during it's evolution back in time unlike in the standard $\Lambda CDM$ one. This is a ``bounce'' of the scale factor from it's minimal value ${a_{m}} \ne 0$. The bounce appears in case there is a local minimum of the scale factor greater than zero. The parameter phase-space for the bounce case starts from ${a_{m}(E)=0}$, so, it is possible to obtain the condition for a bounce from the equation (\ref{Fridman7-5}). It has the following form:
\begin{eqnarray}
\Delta > 0 \, .
\label{B-1}
\end{eqnarray}
Further, the time estimation for $E$ at the bounce is:
\begin{eqnarray}
\quad E_{min}=\sqrt{c^- /c^+}\, .
\label{B-1}
\end{eqnarray}
The exact equality ${E_{m}= \sqrt{c^- /c^+}}$ is satisfied when ${\Delta = 0}$, e.q. when the scale factor vanishes at the local minimum. Such a scenario excludes the initial singularity and leaves the scale factor regular and continuous everywhere, including during the bounce (see (\ref{Fridman7-4})). 

It is possible to estimate the numerical values of the scale factor $\tilde a(t)$ and $\Delta$ at the bounce from the following arguments. The  cosmological microwave background radiation (CMBR) indicates that the Universe was hot and radiation-dominated at early stages of it's evolution \cite{Novikov2006}. Using the expression for an adiabatic expansion ${a_{hot}/a_0 = 4\cdot 10^{-5}}$ \cite{Zeldovich1975en}  we can obtain the value of $\Delta$ at the bounce for ${|w|>> 1}$ from the equation (\ref{initial-4}), so that:
\begin{eqnarray}
\Delta \approx 2\tilde\rho_0 a_{m}^3/a_0^3 < 2\tilde\rho_0 a_{hot}^3/a_0^3 \approx 2.6\cdot 10^{-14}
\label{B-2}
\end{eqnarray}
This tiny value of $\Delta$ can only be achieved in a nearly flat Universe, i.e. when ${1+q_0-\beta\approx 0}$. This result states that the LLR bound on $w$ remains in agreement with the cosmological one in BD$\Lambda$ model, as well as with the flatness of the Universe.

The model under consideration, as well as the usual singular cosmology\footnote{The article ~\cite{Uehara1982} presents an
analytical solution for ${\Delta <0}$, ${\, w>0}$.} of~\cite{Uehara1982} is not applicable for the hot stage of the Universe. Thus the discussed results should only be used to obtain initial values for the hot Universe during the evolution back in
time study.

\section{Solution for a hot Universe}
\label{s6}

The analytical study of the functions behavior near the bounce can be done only for an ultrarelativistic pressure. During the hot phase when ${\,\, p =\frac{1}{3}\rho\,\,}$ equations  (\ref{Fridman1}-\ref{Fridman3}) lead to the following expression:
\begin{eqnarray}
\dot{\tilde H} + 2{\tilde H}^2 = \frac{1}{6}\left(-w\epsilon^2 +\frac{6+8w}{3+2w} \right) \equiv Q(\tilde t) .
\label{hot-1}
\end{eqnarray}
When ${w<-1.5}$, we obtain positive value of $Q$.

In the FRW case (when ${\epsilon =0}$ and ${|w|\to\infty}$), from equation (\ref{hot-1}) it is possible to obtain expressions for the hot stage that are similar to Eqs (\ref{Fridman7-8-1}-\ref{Fridman7-8-2}):
\begin{eqnarray}
H_{Fr}=\frac{1}{\sqrt{3}}\cdot\frac{U+1}{U-1}\, ,\quad
U(\tilde\tau)\equiv\exp\left(\frac{4\tilde\tau}{\sqrt{3}}\right) \, , \quad
\frac{a_{Fr}}{a_{hot}} = \left[\frac{(U-1)^2 U_{hot}}{(U_{hot}-1)^2 U}\right]^{1/4} .
\label{hot-2}\end{eqnarray}
Here a new variable ${\tilde\tau}$ is introduced. It represents a dimensionless time, that is a measure from the scale factor minimum $(\tilde \tau- \tilde t)/\sqrt{\Lambda}$ is equal to the age of the Universe and the subscript ``$hot$'' corresponds to the transition from the hot stage to the cold one.

The derivatives of the scale factor (\ref{hot-2}) are singular (as well as for the FRW case in the matter-dominated Universe --- see~(\ref{Fridman7-8-1}-\ref{Fridman7-8-2})). Remarkably, when ${\Delta\to 0}$ is in the cold phase, the second derivative ${\ddot a}$ goes to ${+\infty}$ at the bounce while it goes to ${-\infty}$ in the vicinity of the bounce. Therefore the Hubble function appears to be rapidly growing when ${\Delta\to 0}$. The situation has to be similar to the case of the radiation-dominated Universe. Since the hot phase matches large values of ${\tilde H}$, it ends for a short time interval ${\tilde\tau}$. Therefore, when ${|w|>>1 }$ and ${\epsilon<< 1}$ the solution of ${BD\Lambda}$ is nearly indistinguishable from the FRW one (except in the bounce region).

Further, we consider the series expansion of the scale factor ${a(\tilde\tau)}$ against $\tilde\tau$ near the local minimum (bounce). Keeping the terms up to the fourth order, it is possible to obtain:
\begin{eqnarray}
a = a_m + \frac{1}{2} a_m \dot{\tilde H}_m {\tilde\tau}^2 - \frac{1}{12} a_m b^2 {\dot{\tilde H}_m^2} {\tilde\tau}^4 + ... .
\label{hot-4}\end{eqnarray}
Here, ${\dot{\tilde H}_m}$ and $b$ are constants; ${\dot{\tilde H}_m}$ corresponds to the second derivative of the scale factor at the bounce, hence it should be positive when ${a_m>0}$: ${\quad \dot{\tilde H}_m >0}$. Therefore, the equations for the Hubble function and it's first derivative up to second order on ${\tilde\tau}$ are:
\begin{eqnarray}
{\tilde H}^2 = {\dot{\tilde H}_m^2} {\tilde\tau}^2 \, ,\quad
\dot{\tilde H} = \frac{\dot{\tilde H}_m (1-b^2 \dot{\tilde H}_m \tilde\tau^2)}{1+\dot{\tilde H}_m \tilde\tau^2 /2} - \dot{\tilde H}_m^2 \tilde\tau^2 \, .
\label{hot-5}\end{eqnarray}
After substituting this into (\ref{hot-1}), one gets:
\begin{eqnarray}
\frac{\ddot a}{a}+\frac{{\dot a}^2}{a^2} =
\frac{\dot{\tilde H}_m\left[ 1 + {\tilde\tau}^2 \dot{\tilde H}_m \left(\frac{3}{2}-b^2\right)\right]}{(1+\dot{\tilde H}_m \tilde\tau^2 /2)^2} = Q > 0
\label{hot-6}\end{eqnarray}
The last inequality is satisfied automatically (see equation (\ref{hot-1})) and is valid only when ${0< b^2< 3/2}$.

From (\ref{hot-4}), one notices that at ${\tilde\tau_1 = 1/\sqrt{\dot{\tilde H}_m b^2}}$ the second derivative of the scale factor changes its sign (${\tilde\tau_1}$ is an inflection point). So we consider an additional scale factor inflection point compared to the FRW case. At the time ${\tilde\tau_2 = \sqrt{3}/\sqrt{\dot{\tilde H}_m b^2} = \sqrt{3}\tilde\tau_1}$, the first derivative of the scale factor changes its sign. Hence, the solutions for the hot phase and the cold one should be matched along the $\tilde\tau_1 $ to $\tilde\tau_2$ interval.

When $\dot{\tilde H}_m$ is large\footnote{Here and below we compare with the unit value} and $b$ is small, starting from the time ${\tilde\tau_1}$, the second derivative of the scale factor rapidly goes to a large negative value (during the time interval of order of ${\tilde\tau_1}$). Meanwhile the Hubble function remains positive (up to the moment ${\tilde\tau_2}$).
Therefore, along the ${\tilde\tau_1}$ to ${\tilde\tau_2}$ interval the solution for the hot phase can be matched to the cold phase solution. Varying the values of $a_m$, ${\dot{\tilde H}_m}$ and $b$, one could achieve a smooth connection.

\section{Conclusions}
\label{s7}

In this paper, we have demonstrated that the Friedman solution with a cosmological term is a degenerated case of a more generic cosmology (for example, the ${BD\Lambda}$ one as a ground effective approximation).

In standard FRW cosmology, the graph of the solution of the scale factor ${a(t)}$ has a form of a vertical line before vanishing (the bounce is possible afterwards); 
in the ${BD\Lambda}$ case 
\begin{itemize}
\item 
with ${w>0}$ the graph of the scale factor ${a(t)}$ vanishes with a finit slope (first derivative remains finite at ${a=0}$); \\
\item
when ${w<0}$ the graph of the scale factor ${a(t)}$ does not reach zero (and a bounce occurs), so all functions remain regular.
\end{itemize}

Further, an adequate model with a bounce can be obtained numerically because of the complicated structure of the theory. Here it is importantly to note that the parameter $k$ (we considered the case $k=0$ in this paper) describes the flatness type of the Universe and could provide a leading contribution near the bounce due to the the scale factor smallness. The hot phase also implies the presence of a non-vanishing pressure (we considered $p=0$) which leads to the inability of obtaining an analytical solution for the hot phase. The presence of the bounce in the BD$\Lambda$ cosmological solution allows to avoid one of the greatest problem of cosmology: the initial singularity.

Finally, we would like to point out that the appearance of a bounce instead of a singularity is ratter a common effect in gravity models when an additional scalar or tensor contribution is taken into account. For example, when studying the interplay between curvature and Maxwell terms in Gauss-Bonnet gravity one encounters an effect of the same nature when the singularity is changed by a local minimum \cite{Alexeyev:2009kx}. The same effects occur in Gauss-Bonnet cosmology with additional fields \cite{sa_cqg}. A bounce appearance was also discovered in many new (sometimes exotic) models with additional terms (some examples of bounce appearance can be found at \cite{Cai:2008qw,Cai:2007qw, Cai:2009in,Cai:2011tc}.)  and in Loop Quantum Gravity \cite{Grain:2010yv,Ashtekar:2011ni}. So, this effect is rather common and natural from the mathematical point of view (changing the balance between different term contributions) and can be used to obtain new (stronger) estimations of model parameters. So, based on the condition of existence of a bounce and, so, from Eq. (\ref{B-2}) one can put a new limit on the BD parameter $w$ in the form
\begin{eqnarray}
|w| > 10^{40}, \qquad w<0. 
\end{eqnarray}
This limitation is much stronger than the existing experimental one ($|w|>50000$) and future developments will show the connection of these values with reality.

\section*{Acknowledgements}
\label{s8}

This work was supported in part by the Federal Program ``Scientific-Pedagogical Innovational Russia 2009-2011"\,
and the program by the presidium of RAS  "The origin, structure and evolution of the universe 2011".  The work was also supported by Federal Agency on Science and Innovations of Russian Federation, state contract 02.740.11.0575.

\end{document}